\documentclass{aa} 

\usepackage{txfonts,natbib,graphicx}

\newcommand{\xt}[1]{\mbox{$\times 10^{#1}$}}

\defcitealias{vandesteene00}{VdS00}
\defcitealias{garcialario97}{GL97}

\begin{document}

   \title{IRAS\,08281-4850 and IRAS\,14325-6428:\\
          two A-type post-AGB stars with s-process enrichment\,\thanks{based on
          observations collected at the European Southern Observatory, Chile
          (programmes 70.D-0278(A) and 73.D-0241(A))}
          }

   \subtitle{}

   \author{M. Reyniers\inst{1}\fnmsep\thanks{Postdoctoral fellow of the Fund for
   Scientific Research, Flanders}
   \and
   G.C. Van de Steene\inst{2}
   \and
   P.A.M. van Hoof\inst{2}
   \and
   H. Van Winckel\inst{1}
   }

   \offprints{M. Reyniers}

   \institute{Instituut voor Sterrenkunde, Departement Natuurkunde en
              Sterrenkunde, K.U.Leuven, Celestijnenlaan 200D, 3001 Leuven,
              Belgium
         \and
              Koninklijke Sterrenwacht van Belgi\"e, Ringlaan 3, 1180 Brussel,
              Belgium}

   \date{Received 4 April 2007 / Accepted 6 June 2007}

   \authorrunning{M. Reyniers et al.}
   \titlerunning{Abundance analysis of IRAS\,08281-4850 and IRAS\,14325-6428}

 
  \abstract
  {} 
   {One of the puzzling findings in the study of the chemical evolution
  of (post-)AGB stars is why very similar stars (in terms of metallicity,
  spectral type, infrared properties, etc\ldots) show a very different
  photospheric composition. We aim at extending the still limited
  sample of s-process enriched post-AGB stars, in order to obtain a
  statistically large enough sample that allows us to formulate conclusions
  concerning the 3rd dredge-up occurrence.}
   {We selected two post-AGB stars on the basis of IR colours indicative
  of a past history of heavy mass loss:  \object{IRAS\,08281-4850} and
  \object{IRAS\,14325-6428}. They are cool sources in the locus of the
  Planetary Nebulae (PNe) in the IRAS colour-colour diagram. Abundances of
  both objects were derived for the first time on the basis of high-quality
  UVES and EMMI spectra, using a critically compiled line list with accurate
  log(gf) values, together with the latest Kurucz model atmospheres.}
   {Both objects have very similar spectroscopically defined effective
  temperatures of 7750\,-\,8000\,K. They are strongly carbon and s-process
  enriched, with a C/O ratio of 1.9 and 1.6, and an [ls/Fe] of $+$1.7 and
  $+$1.2, for \object{IRAS\,08281-4850} and \object{IRAS\,14325-6428}
  respectively. Moreover, the spectral energy distributions (SEDs) point to
  heavy mass-loss during the preceding AGB phase.}
   {\object{IRAS\,08281-4850} and \object{IRAS\,14325-6428} are prototypical
  post-AGB objects in the sense that they show strong post 3rd
  dredge-up chemical enrichments. The neutron irradiation has been
  extremely efficient, despite the only mild sub-solar metallicity.
  This is not conform with the recent chemical models.
  The existence of very similar post-AGB stars
  without any enrichment emphasizes our poor knowledge of the details of the AGB
  nucleosynthesis and dredge-up phenomena. We call for a very
  systematic chemical study of all cool sources in the PN region of
  the IRAS colour-colour diagram.  }


   \maketitle


\section{Introduction}\label{sect:introdctn}
Post-AGB stars are key objects in the study of the dramatic chemical and
morphological changes of objects on their ascent on the Asymptotic Giant
Branch (AGB) and subsequent evolution. In this paper we report on our ongoing
research to study the AGB chemical evolution by a systematic study of post-AGB
photospheres.
Spectra of post-AGB stars are much easier to study than their AGB
precursors for several reasons. First, their atmospheres do not show the large
amplitude pulsations and the large mass loss rates that characterise AGB atmospheres.
Second, their photospheres are hotter, so {\em atomic} transitions prevail in
post-AGB spectra, while molecular transitions prevail in AGB star spectra.
This allows us to quantify the chemical content of a very wide
range of trace elements. Unfortunately,
post-AGB stars evolve on a very fast track and whether
the current Galactic sample is representative is one of the most difficult
issues when interpreting the results of single post-AGB stars in the
broader context of stellar evolution. Therefore, extending
the still limited sample of post-AGB stars is indispensable in order to attain
a statistically large enough sample that allows us to formulate conclusions
concerning the (post-)AGB evolution in general.

During the past decennia, it has been realised that post-AGB stars are
chemically much more diverse than previously thought. {\em Binary objects}
tend to have a totally different photospheric composition than single objects,
showing some degree of {\em depletion} of refractory elements in their
photosphere \citep[see e.g.][]{maas05}. The {\em single} objects, on their
side, are also far from chemically homogeneous. Some objects are the most
s-process enriched objects known to date
\cite[e.g.][and references therein]{reyniers04}, while others are not enriched
at all. This dichotomy is very strict, in the sense that mildly enhanced
objects do not exist, except for a few rather atypical objects.
Chemical evolutionary AGB models do predict that there is a minimum initial
mass for the 3rd dredge-up to occur, at around 1.4\,M$_{\odot}$
\citep[e.g.][]{straniero03}. Hence post-AGB stars without any helium burning
products in their photosphere are theoretically expected. 
However a strict dichotomy is not expected, since a more gradual 
transition between non-enriched and enriched objects is predicted if 
the transition from an O-rich AGB to a C-rich AGB star occurs over
many thermal pulses.
Furthermore, the s-process enriched sub-class of post-AGB
objects imposes another unsolved problem. These stars exhibit a large
spread in s-process efficiency, as they do not obey the expected
anti-correlation between metallicity and s-process efficiency
\citep[see Fig. 4 of][]{vanwinckel03}. In other words, examples exist
of very similar post-AGB stars  (in metallicity, spectral type, infrared
excess, etc\ldots), but with a totally different photospheric abundance pattern
\citep[see for example Fig. 1 in][]{vanwinckel01}. This result dramatically
illustrates that the 3rd dredge-up phenomenon is not yet fully understood.

\begin{table}
\caption{Basic parameters of the two objects discussed in this study.}\label{tab:basicprmtrs}
\begin{center}
\begin{tabular}{rrrr}
\hline\hline
IRAS         & & 08281-4850  & 14325-6428 \\ 
\hline
Equatorial coord. & $\alpha_{2000}$ &   08 29 40.552 &   14 36 34.375 \\
                  & $\delta_{2000}$ & $-$49 00 04.33 & $-$64 41 31.09 \\
Galactic coord. & l           &         266.08   &      313.87 \\
                & b           &        $-$5.82   &     $-$4.08 \\
Vis. magnitude  & V           &           14.1   &        11.9 \\
Spectral Type   &             & A9I              &         A8I \\
\hline
\multicolumn{4}{l}{\small Source SIMBAD, except magnitude and spectral type}\\
\multicolumn{4}{l}{\small (this work).}
\end{tabular}
\end{center}
\end{table}

In this paper, we selected two objects with far infrared colours typical
of PNe from the IRAS point source catalogue. Apart from PNe, only post-AGB
stars are typically found in this part of the colour-colour diagram
\citep{vanhoof97}. Furthermore, the selected objects were not detected in the
radio continuum above a detection limit of 3~mJy
\citep{vandesteene93, vandesteene95}. Hence they do not seem to have evolved to
the PN stage as yet and are very likely post-AGB stars. We obtained
JHKL images of the candidates with CASPIR on the 2.3-m telescope at Siding
Spring Observatory in Australia in order to assure the correct identification
of the IRAS counterparts and obtain accurate positions \citep{vandesteene00}.
Two of these infrared selected post-AGB stars which are presented here (see
Table \ref{tab:basicprmtrs}) have counterparts in the USNO
catalogue so that their visual magnitudes are also known. They were sufficiently bright to be observed in the optical at high resolution.

The paper is organised as follows: in the next section, we discuss
the spectral energy distribution of the two objects, and quantify the
total reddening towards the two sources. In Sect. \ref{sect:observtns}
we briefly discuss the observation and reduction of the
high-resolution spectra, while in Sect. \ref{sect:analysisrslts} we
deal with the abundance analysis.  Sect. \ref{sect:diffuseisbnds} is a
section devoted to the diffuse interstellar bands in the spectrum of
\object{IRAS\,14325-6428}. In the discussion
(Sect. \ref{sect:discssn}), we mainly focus on the s-process
abundances.  We end the paper with conclusions.


\section{Spectral energy distribution}

\subsection{Geneva Photometry}
Images were obtained for both objects with the Swiss Euler telescope
at La Silla (ESO) in 2005. The observing log is presented in Table~\ref{obsphot}.
The objects were unresolved in all images.
The images were bias subtracted and flatfielded in {\sc iraf}.  The photometry
package was used to do the aperture photometry. In case of crowded fields
{\sc daophot} was used to subtract the neighbours.  One to three standard stars
per band per star were available for the photometric calibration.  The
resulting magnitudes are presented in Table~\ref{addphot}.

\begin{table}
\caption{Observational log for the images obtained at the Euler telescope at
ESO, La Silla.}\label{obsphot}
\begin{center}
\begin{tabular}{rrrrr}
\hline\hline
IRAS & Date &  filter & exptime & airmass \\
\hline
08281-4850 & 2005-10-07 & UG & 300.2 & 1.473 \\
             &            & BG & 20.9  & 1.423 \\
             &            & VG & 150.5 & 1.408 \\
             &            & RG & 60.6  & 1.392 \\
             &            & IC & 98.9  & 1.351 \\
\hline
14325-6428 & 2005-08-04 & UG & 399.3 & 1.322 \\
             &            & BG & 299.5 & 1.311 \\ 
             &            & VG & 241.6 & 1.292 \\
\hline
\end{tabular}
\end{center}
\end{table}


\subsection{Additional Photometry}

We collected additional photometry from various surveys: the Tycho-2 Catalogue
\citep{hog00}, the 3$^{\rm rd}$ data release of DENIS \citep{epchtein94}, 2MASS
\citep{skrutskie97}, MSX \citep{egan03}, and the IRAS point source catalogue
\citep{beichmann85}. We also obtained literature data from
\citet[][hereafter GL97]{garcialario97} and \citet[][hereafter VdS00]{vandesteene00}.
The data are presented in Table~\ref{addphot}. Near-IR data
of \object{IRAS\,08281-4850} were present in the 2$^{\rm nd}$ release of the
DENIS data set, but was deleted in the 3$^{\rm rd}$ release. The data from the
2$^{\rm nd}$ release have not been retained in the analysis. In order to
construct the spectral energy distribution (SED), we first converted the
Tycho-2 magnitudes to the Johnson system using the formulas:
\[ {\rm B}_J = {\rm B}_T - 0.240({\rm B}_T - {\rm V}_T) \]
\[ {\rm V}_J = {\rm V}_T - 0.090({\rm B}_T - {\rm V}_T) \]
which were derived from \citet{perryman97}. The Gunn-r magnitude from
the Geneva system was converted to the Johnson system using the formula:
\[ {\rm R}_J = {\rm r} - 0.43 - 0.15({\rm B}_J - {\rm V}_J) \]
\citep{kent85} where B$_J$ $-$ V$_J$ can be calculated from the Geneva
photometry using Eqs.~12 and 13 from \citet{harmanec01}. After that, the
photometry was converted to $\lambda F_\lambda$ using the absolute flux
calibrations given in Table~\ref{absflux}. For the \citetalias{garcialario97}
observations we adopt the Johnson photometric calibration, while for the
\citetalias{vandesteene00} observations we adopt the MSSO photometric
calibration. The resulting spectral energy distribution is shown in
Table~\ref{nufnu}.

\begin{table}
\caption{Broadband photometry for \object{IRAS\,08281-4850} and
\object{IRAS\,14325-6428}. \object{IRAS\,14325-6428} is contained twice in
the DENIS catalogue. The data sets do not agree within the error margins, which
may point to variability of the source. Both data sets have been retained in
the analysis.}\label{addphot}
\begin{center}
\begin{tabular}{lrrrr}
\hline\hline
phot.\ band & \multicolumn{2}{c}{08281-4850} & \multicolumn{2}{c}{14325-6428} \\
            & mag & $\Delta$mag & mag & $\Delta$mag \\
\hline
Euler Geneva U & 17.81 & 0.15 & 14.58 & 0.25 \\
Euler Geneva B & 14.76 & 0.10 & 12.08 & 0.10 \\
Euler Geneva V & 14.09 & 0.05 & 11.89 & 0.05 \\
Euler Gunn-r   & 13.23 & 0.15 &       &      \\
Euler I$_C$    & 12.20 & 0.05 &       &      \\
Tycho-2 B$_T$  &       &      & 12.676& 0.209\\
Tycho-2 V$_T$  &       &      & 12.065& 0.169\\
DENIS Gunn-i   &       &      & 10.297& 0.04 \\
DENIS J        &       &      &  9.202& 0.07 \\
DENIS K$_s$    &       &      &  8.445& 0.09 \\
DENIS Gunn-i   &       &      & 10.431& 0.02 \\
DENIS J        &       &      &  9.292& 0.05 \\
DENIS K$_s$    &       &      &  8.604& 0.05 \\
2MASS J        & 10.603& 0.019&  9.263& 0.023\\
2MASS H        & 10.124& 0.023&  8.860& 0.025\\
2MASS K$_s$    &  9.835& 0.021&  8.598& 0.025\\
\citetalias{garcialario97} J & 10.68 & 0.05 &       &      \\ 
\citetalias{garcialario97} H & 10.12 & 0.05 &       &      \\ 
\citetalias{garcialario97} K &  9.79 & 0.04 &       &      \\ 
\citetalias{vandesteene00} J&&&  9.27 & 0.05 \\
\citetalias{vandesteene00} H&&&  8.81 & 0.05 \\
\citetalias{vandesteene00} K&&&  8.61 & 0.05 \\
\citetalias{vandesteene00} L&&&  8.27 & 0.10 \\
\hline
            & F$_\nu$ & $\Delta$F$_\nu$ & F$_\nu$ & $\Delta$F$_\nu$ \\
            & (Jy) & (Jy) & (Jy) & (Jy) \\
\hline
MSX A 8.28     &       &      &  0.72 & 0.03 \\
MSX C 12.13    &       &      &  3.35 & 0.18 \\
MSX D 14.65    &       &      &  7.1  & 0.4  \\
MSX E 21.34    &       &      & 17.7  & 1.1  \\
IRAS 12        &  2.23 & 0.11 &  3.31 & 0.20 \\
IRAS 25        &  9.83 & 0.59 & 30.6  & 1.2  \\
IRAS 60        &  3.62 & 0.36 & 17.1  & 1.9  \\
\hline
\end{tabular}
\end{center}
\end{table}

\onltab{4}{
\begin{table}
\caption{Absolute flux calibration for the various photometric bands used in
this paper.}\label{absflux}
\begin{tabular}{lcc}
\hline\hline
phot.\ band & 0-mag flux & reference \\
            & W\,m$^{-2}$\,$\mu$m$^{-1}$ \\
\hline
2MASS J      & 3.129\xt{-9}  & \citealt{cohen03} \\
2MASS H      & 1.133\xt{-9}  &  '' \\
2MASS K$_s$  & 4.283\xt{-10} &  '' \\
Cousins I$_C$& 1.196\xt{-8}  & \citealt{lamla82} \\
DENIS Gunn-i & 1.20\xt{-8}   & \citealt{fouque00} \\
DENIS J      & 3.17\xt{-9}   &  '' \\
DENIS K$_s$  & 4.34\xt{-10}  &  '' \\
Geneva U     & 5.754\xt{-8}  & \citealt{rufener88} \\
Geneva B     & 2.884\xt{-8}  &  '' \\
Geneva V     & 3.736\xt{-8}  &  '' \\
Johnson B    & 7.20\xt{-8}   & \citealt{johnson65, johnson66} \\
Johnson V    & 3.92\xt{-8}   &  '' \\
Johnson R    & 1.76\xt{-8}   &  '' \\
Johnson J    & 3.4\xt{-9}    &  '' \\
Johnson H    & 1.26\xt{-9}   &  '' \\
Johnson K    & 3.9\xt{-10}   &  '' \\
MSSO J       & 3.03\xt{-9}   & \citealt{thomas73} \\
MSSO H       & 1.17\xt{-9}   &  '' \\
MSSO K       & 4.02\xt{-10}  &  '' \\
MSSO L       & 6.18\xt{-11}  &  '' \\
\hline
\end{tabular}
\end{table}}

\onltab{5}{
\begin{table}
\caption{The spectral energy distribution (sorted by wavelength) given as
$\lambda F_\lambda$ in SI units.}\label{nufnu}
\begin{tabular}{rcccc}
\hline\hline
$\lambda$ & \multicolumn{2}{c}{08281-4850} & \multicolumn{2}{c}{14325-6428} \\
          & $\lambda F_\lambda$ & $\Delta\lambda F_\lambda$ & $\lambda F_\lambda$ & $\Delta\lambda F_\lambda$ \\
$\mu$m    & W\,m$^{-2}$  & W\,m$^{-2}$  & W\,m$^{-2}$  & W\,m$^{-2}$ \\
\hline
 0.3464 & 1.50\xt{-15} & 2.2\xt{-16} & 2.94\xt{-14} & 7.6\xt{-15} \\
 0.4227 & 1.52\xt{-14} & 1.5\xt{-15} & 1.79\xt{-13} & 1.7\xt{-14} \\
 0.4442 &              &             & 3.11\xt{-13} & 6.6\xt{-14} \\
 0.5488 & 4.74\xt{-14} & 2.2\xt{-15} & 3.60\xt{-13} & 1.7\xt{-14} \\
 0.5537 &              &             & 3.41\xt{-13} & 5.8\xt{-14} \\
 0.6938 & 1.10\xt{-13} & 1.6\xt{-14} &              &             \\
 0.7886 & 1.24\xt{-13} & 5.9\xt{-15} &              &             \\
 0.7910 &              &             & 6.38\xt{-13} & 1.2\xt{-14} \\
 0.7910 &              &             & 7.22\xt{-13} & 2.7\xt{-14} \\
 1.2280 &              &             & 7.47\xt{-13} & 3.5\xt{-14} \\
 1.2280 &              &             & 8.12\xt{-13} & 5.4\xt{-14} \\
 1.2350 & 2.22\xt{-13} & 5.5\xt{-15} & 7.62\xt{-13} & 2.1\xt{-14} \\
 1.2500 & 2.27\xt{-13} & 1.1\xt{-14} & 7.42\xt{-13} & 3.5\xt{-14} \\
 1.6200 & 1.83\xt{-13} & 8.6\xt{-15} &              &             \\
 1.6500 &              &             & 5.78\xt{-13} & 2.7\xt{-14} \\
 1.6620 & 1.68\xt{-13} & 4.9\xt{-15} & 5.38\xt{-13} & 1.6\xt{-14} \\
 2.1450 &              &             & 3.37\xt{-13} & 1.6\xt{-14} \\
 2.1450 &              &             & 3.90\xt{-13} & 3.4\xt{-14} \\
 2.1590 & 1.08\xt{-13} & 2.9\xt{-15} & 3.36\xt{-13} & 1.0\xt{-14} \\
 2.1900 & 1.04\xt{-13} & 3.9\xt{-15} &              &             \\
 2.2000 &              &             & 3.18\xt{-13} & 1.5\xt{-14} \\
 3.6000 &              &             & 1.09\xt{-13} & 1.1\xt{-14} \\
 8.2800 &              &             & 2.59\xt{-13} & 1.1\xt{-14} \\
12.0000 & 5.57\xt{-13} & 2.7\xt{-14} & 8.27\xt{-13} & 5.0\xt{-14} \\
12.1300 &              &             & 8.27\xt{-13} & 4.4\xt{-14} \\
14.6500 &              &             & 1.45\xt{-12} & 8.8\xt{-14} \\
21.3400 &              &             & 2.48\xt{-12} & 1.5\xt{-13} \\
25.0000 & 1.18\xt{-12} & 7.1\xt{-14} & 3.67\xt{-12} & 1.4\xt{-13} \\
60.0000 & 1.81\xt{-13} & 1.8\xt{-14} & 8.54\xt{-13} & 9.5\xt{-14} \\
\hline
\end{tabular}
\end{table}}

\subsection{Fitting the SED}
\begin{figure}
\resizebox*{\hsize}{!}{\includegraphics{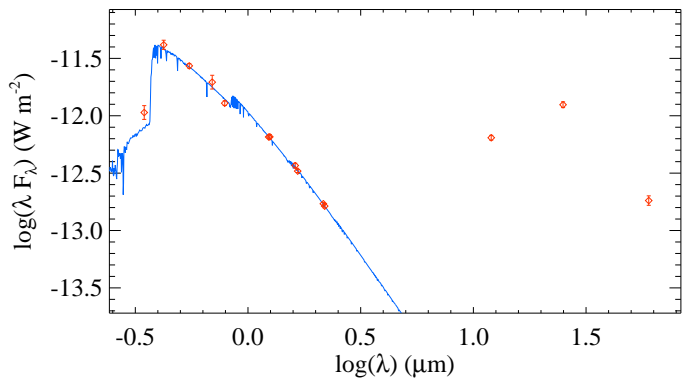}}
\caption{The spectral energy distribution of \object{IRAS\,08281-4850}.\label{sedi}}
\end{figure}

\begin{figure}
\resizebox*{\hsize}{!}{\includegraphics{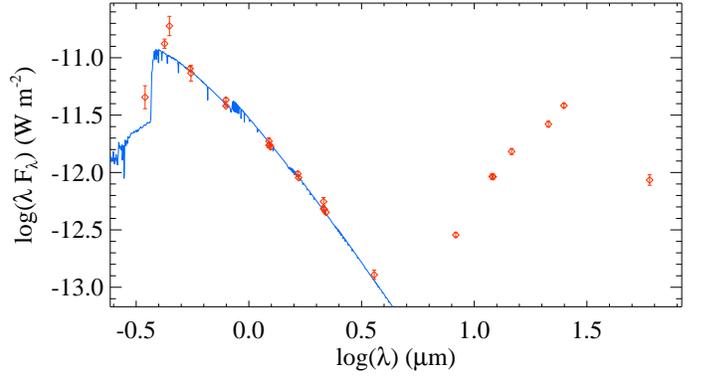}}
\caption{The spectral energy distribution of \object{IRAS\,14325-6428}.\label{sedii}}
\end{figure}

The SEDs of the post-AGB stars suffer from interstellar as well as
circumstellar extinction. We assumed that this total extinction can be
described by an $R_V$\,=\,3.1 galactic extinction curve as defined by
\citet{fitzpatrick99}. We furthermore assumed that the intrinsic spectrum of
the stars can be described by an Atlas stellar atmosphere model
\citep{castelli04} with the parameters given in Table~\ref{tab:abndcsboth}.
We then used an iterative procedure to determine the amount of total
extinction $A_V$ and the bolometric flux $L/(4\pi D^2)$ by minimizing the
quadratic residuals of all observations between 0.4 and 4~$\mu$m. The results
are given in Table~\ref{tab:sedsfinalrslts}. Plots of the Kurucz stellar
atmosphere models combined with the dereddened observed fluxes are shown in
Figs.~\ref{sedi} and \ref{sedii}.

\begin{table}
\caption{The total extinction $A_V$ and the bolometric luminosity $L$
of the program stars.}\label{tab:sedsfinalrslts}
\begin{center}
\begin{tabular}{lrr}
\hline\hline
IRAS & $A_V$ & $L$ at D = 5~kpc \\
     & mag & L$_\odot$ \\
\hline
08281-4850 & 4.33 & 2450 \\
14325-6428 & 3.32 & 7250 \\
\hline
\end{tabular}
\end{center}
\end{table}

\section{High-resolution spectra: observation and reduction}\label{sect:observtns}
High-resolution, high signal-to-noise optical spectra of the two program
stars were taken in the framework of our ongoing program to study the
photospheric chemical composition of stars in their last stages of evolution
\citep[e.g.][]{reyniers04, reyniers03,vandesteene03}. \object{IRAS\,14325-6428}
is observed with UVES on the VLT UT2 telescope (Kueyen), as a member of a
larger sample of seven post-AGB objects that were observed in service mode
during ESO period \#73. \object{IRAS\,08281-4850} was observed in visitor mode
with the EMMI echelle grating \#9 with cross disperser \#3 at the
NTT during period \#70. The resolving power of the UVES spectra varies between
$\sim$55,000 and $\sim$60,000. The resolving power of our EMMI spectra is
significantly lower ($\sim$8,400). The spectral interval covered and some other
details about the observations are given in Table~\ref{tab:observations}.

\begin{figure}
\resizebox{\hsize}{!}{\includegraphics{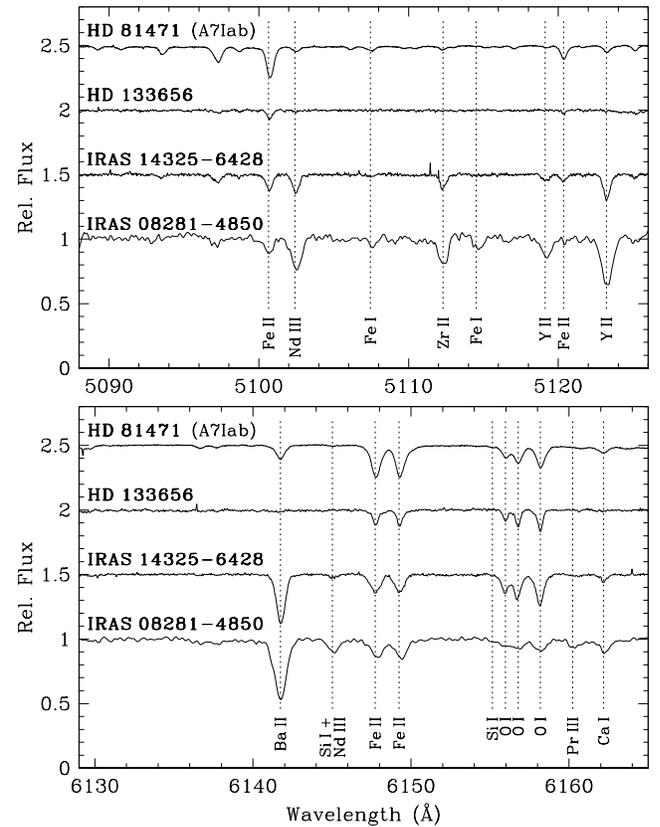}}
\caption{The spectra of \object{IRAS\,14325-6428} and \object{IRAS\,08281-4850}
compared with the A7Iab standard \object{HD\,81471} and the non-enriched
post-AGB star \object{HD\,133656}.}\label{fig:rsmblnce}
\end{figure}

\begin{table}\caption{Log of the high-resolution observations. For
\object{IRAS\,08281-4850}, a spectral gap occurs between 493\,nm and 500\,nm.
For \object{IRAS\,14325-6428}, spectral gaps occur between 577\,nm and 583\,nm
and between 854.4\,nm and 864.5\,nm due to the spatial gap between the two
UVES CCDs.}\label{tab:observations}
\begin{center}
\begin{tabular}{ccccc}
\hline\hline
 date & UT    & exp.time & wavelength & S/N \\
      & start &  (sec)   & interval (nm) &     \\
\hline
\multicolumn{5}{c}{\object{IRAS\,08281-4850} (NTT\,+\,EMMI)}\\
\hline
 2003-02-04 & 01:11 & 9000 & 395$-$795 & 75 \\ 
\hline
\multicolumn{5}{c}{\object{IRAS\,14325-6428} (VLT-UT2\,+\,UVES)}\\
\hline
 2004-05-13 & 07:13 & 1800 & 374.5$-$498 & 130 \\ 
 2004-05-13 & 07:13 & 1800 & 670.5$-$1055 & 170 \\ 
 2004-05-13 & 06:39 & 1800 & 477.5$-$681 & 190 \\ 
\hline
\end{tabular}
\end{center}
\end{table}

The reduction of our UVES spectra was performed in the dedicated
``UVES context'' of the {\sc midas} environment and included bias correction,
cosmic hit correction, flat-fielding, background correction and sky correction.
We used average extraction to convert frames from pixel-pixel to pixel-order
space. The reduction of our EMMI spectrum was done within the echelle package
in {\sc iraf} following the user's guide by D. Willmarth and J. Barnes (1994).
The spectra were normalised by dividing the individual orders by a smoothed
spline function defined through interactively identified continuum points. For
a detailed description of the reduction procedure, we refer to
\citet{reyniers02a}. In Table~\ref{tab:observations}, we also list some
indicative signal-to-noise values of the final data product.

Sample spectra of our programme stars can be found in Figs.~\ref{fig:rsmblnce}
and \ref{fig:fourdbs}. In Fig.~\ref{fig:rsmblnce}, the spectra of
\object{IRAS\,08281-4850} and \object{IRAS\,14325-6428} are compared
with the spectra of the A7Iab standard \object{HD\,81471} and the
non-enriched post-AGB star \object{HD\,133656}. The spectrum of
\object{HD\,81471} is retrieved from UVESPOP, the library of
high-resolution UVES spectra of stars across the Hertzsprung-Russell
diagram \citep{bagnulo03}. \object{HD\,133656} is discussed in
\citet{vanwinckel96}.  The spectrum of \object{HD\,133656} is an
ESO1.5-m+FEROS spectrum taken on June 27, 2001.  This object has
atmospheric parameters comparable with the program stars (T$_{\rm
eff}$, $\log g$, $\xi_t$) = (8000\,K, 1.0, 3.0\,km\,s$^{-1}$) but is
slightly more metal deficient ([Fe/H]\,=\,$-$0.9). The s-process
enhancement of the two IRAS stars is clear from the Ba\,{\sc ii} line.
\object{IRAS\,08281-4850} is the strongest s-enriched one, but the stronger
lines are also due to a slightly lower temperature of this object.


\begin{figure}
\resizebox*{\hsize}{!}{\includegraphics{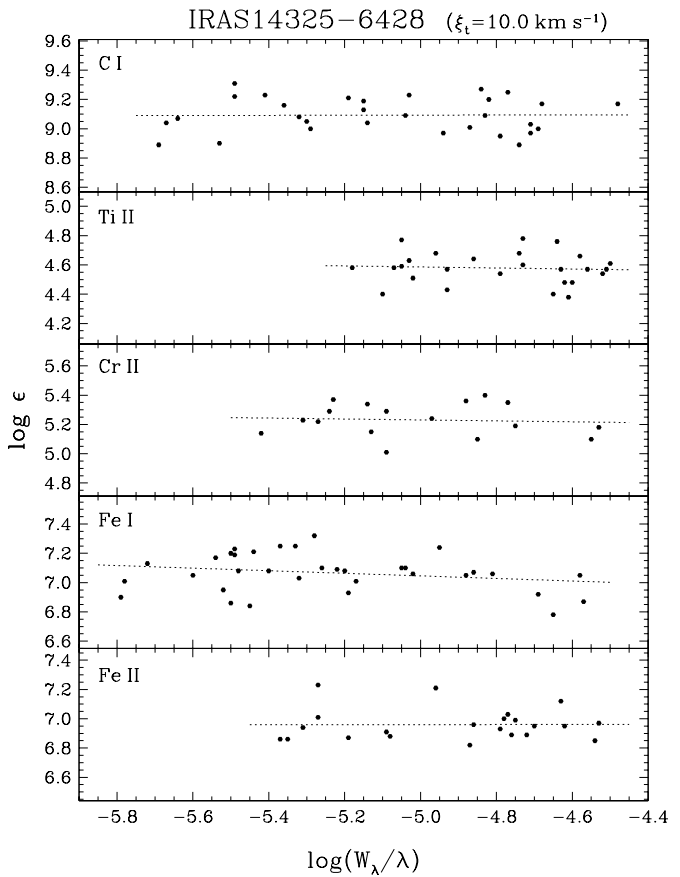}}
\caption{The relatively high microturbulent velocity
($\xi_{\rm t}$\,$=$\,10\,km\,s$^{-1}$) for \object{IRAS\,14325-6428} is also
supported by the results of ions other than Fe\,{\sc ii}.}\label{fig:microtrblnt}
\end{figure}

\section{Abundance analysis and results}\label{sect:analysisrslts}
\subsection{Method}
The general methodology is already extensively discussed in our previous papers
\citep[e.g.][]{reyniers04,deroo05} and will not be repeated here. Here we only
remind that we make use of the latest Atlas models \citep{castelli04} in
combination with the latest version (April 2002) of Sneden's LTE line analysis
program MOOG \citep{sneden73}. The value of the microturbulent velocity of
\object{IRAS\,14325-6428} derived from the Fe\,{\sc ii} lines is high
($\xi_{\rm t}$\,$=$\,10\,km\,s$^{-1}$), but this high value is also supported
by the results of other ions as can be seen in Fig. \ref{fig:microtrblnt}.

Since the spectral resolution of the UVES spectra is much higher than that
of the EMMI spectrum, we started the analysis with
\object{IRAS\,14325-6428}. In order to avoid undetected blends that
are caused by the lower spectral resolution of the EMMI spectrum, we
took the line list of \object{IRAS\,14325-6428} as our starting point
for the analysis of \object{IRAS\,08281-4850}. As a consequence, all
lines that we used for \object{IRAS\,08281-4850} are also present in
the line list of \object{IRAS\,14325-6428}. This choice ensures
the consistency of the analyses, implying that a comparison between
the abundance results is highly reliable. Due to the stronger s-process
enrichment of \object{IRAS\,08281-4850} compared to
\object{IRAS\,14325-6428}, the spectrum of \object{IRAS\,08281-4850} shows
significantly more lines of s-process elements. Therefore, we performed an
extra search for s-process lines in the spectrum of \object{IRAS\,08281-4850}.
This explains why we have used 12 La\,{\sc ii} lines and 1 Nd\,{\sc ii} in
the analysis of \object{IRAS\,08281-4850}, against only 5 La\,{\sc ii} lines
and no Nd\,{\sc ii} lines for \object{IRAS\,14325-6428}.

\begin{table*}\caption{Abundance results for \object{IRAS\,08281-4850} and
\object{IRAS14325-6428}. For the explanation of the columns: see
Sect.~\ref{sect:abundancerslts}.}\label{tab:abndcsboth}
\begin{center}
\begin{tabular}{l|rrrrrr|r|rrrrrr}
\hline\hline
 \multicolumn{1}{c}{}
& \multicolumn{6}{c}{\rule[-0mm]{0mm}{5mm}{\Large\bf IRAS\,08281-4850}}&\multicolumn{1}{c}{}
&\multicolumn{6}{c}{\rule[-0mm]{0mm}{5mm}{\Large\bf IRAS\,14325-6428}}\\
\multicolumn{1}{c}{}
& \multicolumn{6}{c}{
$
\begin{array}{r@{\,=\,}l}
{\rm T}_{\rm eff} & 7750\,{\rm K}  \\
\log g  & 1.0\ {\rm(cgs)} \\
\xi_{\rm t} & 4.5\ {\rm km\,s}^{-1} \\
\end{array}
$
}&\multicolumn{1}{c}{}&
\multicolumn{6}{c}{
$
\begin{array}{r@{\,=\,}l}
{\rm T}_{\rm eff} & 8000\,{\rm K}  \\
\log g  & 1.0\ {\rm(cgs)} \\
\xi_{\rm t} & 10.0\ {\rm km\,s}^{-1} \\
\end{array}
$}\\
\hline
  ion & N &{\rule[0mm]{0mm}{4mm}$\overline{W_{\lambda}}$}&$\log\epsilon$&$\sigma_{\rm ltl}$&[el/Fe]&$\sigma_{\rm tot}$& sun&
        N &{\rule[0mm]{0mm}{4mm}$\overline{W_{\lambda}}$}&$\log\epsilon$&$\sigma_{\rm ltl}$&[el/Fe]&$\sigma_{\rm tot}$\\
\hline
 C\,{\sc i  } &  12 & 117 &  9.17 &  0.24 &  0.93 & 0.15 &  8.57 &  30 &  64 &  9.09 &  0.12 &  1.07 & 0.13\\
 N\,{\sc i  } &   2 &  87 &  8.21 &  0.14 &  0.55 & 0.20 &  7.99 &   6 &  94 &  7.98 &  0.18 &  0.54 & 0.14\\
 O\,{\sc i  } &   2 &  84 &  8.89 &  0.10 &  0.36 & 0.22 &  8.86 &  10 &  72 &  8.89 &  0.11 &  0.58 & 0.14\\
\hline
Na\,{\sc i  } &   1 &  61 &  6.76 &       &  0.76 & 0.25 &  6.33 &   2 &  16 &  6.50 &  0.02 &  0.72 & 0.20\\
Mg\,{\sc i  } &   1 & 107 &  7.31 &       &  0.10 & 0.24 &  7.54 &   3 &  72 &  7.52 &  0.13 &  0.53 & 0.18\\
Mg\,{\sc ii } &     &     &       &       &       &      &  7.54 &   3 &  56 &  7.25 &  0.09 &  0.26 & 0.19\\
Al\,{\sc i  } &     &     &       &       &       &      &  6.47 &   1 & 147 &  5.74 &       &$-$0.18& 0.23\\
 S\,{\sc i  } &     &     &       &       &       &      &  7.33 &   1 &  17 &  7.26 &       &  0.48 & 0.24\\
Ca\,{\sc i  } &   2 &  72 &  6.40 &  0.08 &  0.37 & 0.20 &  6.36 &   2 &  18 &  6.20 &  0.02 &  0.39 & 0.20\\
Sc\,{\sc ii } &   3 & 110 &  3.25 &  0.11 &  0.41 & 0.12 &  3.17 &   4 &  55 &  3.01 &  0.07 &  0.39 & 0.11\\
Ti\,{\sc ii } &   9 & 113 &  4.81 &  0.15 &  0.12 & 0.07 &  5.02 &  26 &  84 &  4.58 &  0.11 &  0.11 & 0.04\\
Cr\,{\sc i  } &   1 & 122 &  5.60 &       &  0.26 & 0.22 &  5.67 &   2 &  61 &  5.34 &  0.10 &  0.22 & 0.16\\
Cr\,{\sc ii } &   9 &  70 &  5.40 &  0.26 &  0.06 & 0.10 &  5.67 &  17 &  58 &  5.23 &  0.11 &  0.11 & 0.04\\
Mn\,{\sc ii } &     &     &       &       &       &      &  5.39 &   2 &  24 &  5.07 &  0.11 &  0.23 & 0.15\\
Fe\,{\sc i  } &  19 &  79 &  7.29 &  0.14 &  0.11 & 0.11 &  7.51 &  34 &  37 &  7.07 &  0.13 &  0.11 & 0.11\\
Fe\,{\sc ii } &  15 &  98 &  7.18 &  0.11 &  0.00 &      &  7.51 &  22 &  74 &  6.96 &  0.11 &  0.00 &     \\
Ni\,{\sc ii } &     &     &       &       &       &      &  6.25 &   3 &  33 &  5.67 &  0.24 &$-$0.03& 0.12\\
\hline
 Y\,{\sc ii } &   9 & 116 &  3.81 &  0.17 &  1.90 & 0.08 &  2.24 &  15 &  70 &  3.01 &  0.19 &  1.32 & 0.06\\
Zr\,{\sc ii } &   1 & 170 &  3.85 &       &  1.58 & 0.26 &  2.60 &  14 &  91 &  3.19 &  0.14 &  1.14 & 0.06\\
\hline
Ba\,{\sc ii } &     &     &       &       &       &      &  2.13 &   1 &  84 &  3.13 &       &  1.55 & 0.20\\

La\,{\sc ii } &  12 &  75 &  2.93 &  0.18 &  2.13 & 0.08 &  1.13 &   5 &  54 &  1.89 &  0.10 &  1.31 & 0.06\\

Ce\,{\sc ii } &     &     &       &       &       &      &  1.58 &   2 &  16 &  2.15 &  0.05 &  1.12 & 0.15\\
Nd\,{\sc ii } &   1 &  52 &  2.78 &       &  1.61 & 0.21 &  1.50 &     &     &       &       &       &	   \\
Nd\,{\sc iii} &   1 & 101 &  2.57 &       &  1.40 & 0.22 &  1.50 &   3 &  60 &  2.00 &  0.11 &  1.05 & 0.14\\
Sm\,{\sc ii } &     &     &       &       &       &      &  1.01 &   1 &  11 &  1.69 &       &  1.23 & 0.20\\
\hline
\multicolumn{1}{c}{}& \multicolumn{7}{c}{summary} &            \multicolumn{6}{c}{summary}\\
\hline
\multicolumn{1}{c}{}& \multicolumn{7}{l}{[Fe/H]\,=\,$-$0.33} & \multicolumn{6}{l}{[Fe/H]\,=\,$-$0.55}\\
\multicolumn{1}{c}{}& \multicolumn{7}{l}{C/O\,=\,1.9} &        \multicolumn{6}{l}{C/O\,=\,1.6}\\
\multicolumn{1}{c}{}& \multicolumn{7}{l}{[$\alpha$/Fe]\,=\,$+$0.2 ($\alpha$: Mg, Ca, Ti)  }& \multicolumn{5}{l}{[$\alpha$/Fe]\,=\,$+$0.3 ($\alpha$: Mg, S, Ca, Ti)}\\ 
\hline
\multicolumn{1}{c}{}& \multicolumn{7}{l}{[ls/Fe]\,=\,$+$1.74} & \multicolumn{6}{l}{[ls/Fe]\,=\,$+$1.23}\\
\multicolumn{1}{c}{}& \multicolumn{7}{l}{[hs/Fe]\,=\,$+$1.91} & \multicolumn{6}{l}{[hs/Fe]\,=\,$+$1.29}\\
\multicolumn{1}{c}{}& \multicolumn{7}{l}{[hs/ls]\,=\,$+$0.17} & \multicolumn{6}{l}{[hs/ls]\,=\,$+$0.06}\\

\hline
\end{tabular}
\end{center}
\end{table*}

\begin{figure}
\resizebox*{\hsize}{!}{\includegraphics{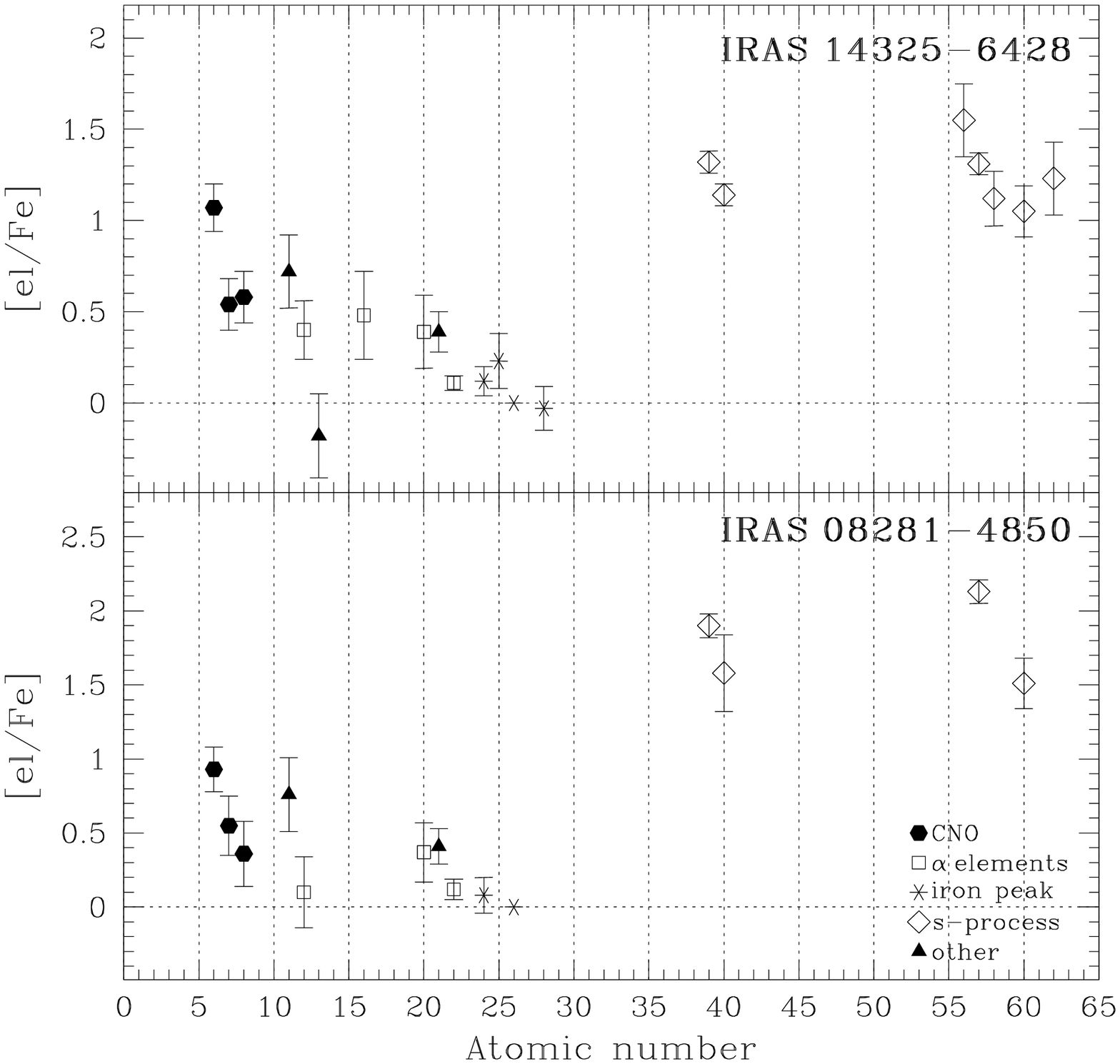}}
\caption{[el/Fe] values for \object{IRAS\,14325-6428} ({\em upper panel}) and
\object{IRAS\,08281-4850} ({\em lower panel}). The uncertainty on the [el/Fe]
values is the total uncertainty $\sigma_{\rm tot}$ as listed in
Table~\ref{tab:abndcsboth}.}\label{fig:abndcsboth}
\end{figure}

\subsection{Abundance results}\label{sect:abundancerslts}
The final results of our stellar atmosphere parameter determination and our
abundance analysis are listed in Table \ref{tab:abndcsboth}. The first column
of this table gives the actual ion. Then, for each star, the following columns
are listed: the number of lines used; the mean equivalent width in m\AA; the
absolute abundances by number $\log\epsilon$\,$=$\,$\log$(N(el)/N(H))+12; the
line-to-line scatter $\sigma_{\rm ltl}$; the abundance relative to iron
[el/Fe], and an estimate of the total uncertainty on the abundance
$\sigma_{\rm tot}$ (see Sect. \ref{subsect:errorestm}). For the references of
the solar abundances (the middle column) needed to calculate the [el/Fe]
values: see \citet{reyniers07a}.

The abundances are also graphically presented in Fig. \ref{fig:abndcsboth}.
On this figure, the different groups of elements are marked with different
symbols. We will summarize the main results for each of these groups.
\begin{description}
\item{\em Metallicity}
Both stars are mildly metal deficient, with iron abundances of
[Fe/H]\,=\,$-$0.33 and $-$0.55 for \object{IRAS\,08281-4850} and
\object{IRAS\,14325-6428} respectively. The other iron peak elements, if
present, follow this deficiency.
\item{\em CNO-elements}
Both stars are clearly carbon enriched, with an enrichment around
[C/Fe]\,$\simeq$\,$+$1 for both stars. As a consequence, we derive also
high C/O number ratios for both stars. One has to note, however, that in
\object{IRAS\,08281-4850} the uncertainty on the oxygen abundance prevents an
accurate C/O number ratio for this star.
\item{\em $\alpha$-elements}
The simple mean of the [el/Fe] values of the (available) $\alpha$-elements
yields [$\alpha$/Fe]=$+$0.2 and $+$0.3 for \object{IRAS\,08281-4850} and
\object{IRAS\,14325-6428} respectively. Such an enhancement is normal for stars
in this metallicity range, as a consequence of the galactic chemical evolution
and therefore does not correspond to an intrinsic enhancement.
\item{\em s-process elements} It is clear that the s-process enrichment
of the two objects under study is very strong.
\end{description}
The s-process elements observed in evolved stars can be divided into two
groups: the light s-process elements around the magic neutron number 50 (Sr,
Y, Zr) and the heavy s-process elements around the magic neutron number 82
(Ba, La, Ce, Pr, Nd, Sm). Three s-process indices are generally defined:
[ls/Fe], [hs/Fe] and [hs/ls]. To be consistent with our earlier papers on
similar stars \citep{vanwinckel00, reyniers04, reyniers05}, we define the
ls-index as the mean of the Y and Zr abundances and the hs-index as the mean
of the Ba, La, Nd and Sm abundances, with unavailable elements estimated using
the tables of \citet{malaney87}. All indices are listed in
Table \ref{tab:abndcsboth}.

\subsection{Uncertainty estimates}\label{subsect:errorestm}
For the error analysis, we followed the same method as described in
\citet{deroo05}. We slightly changed formula (1) of this paper, in the sense
that for the uncertainty induced by the model $\sigma_{\rm mod}$, we confined
the parameter space to {\em consistent} models, i.e. models for which there is
ionisation equilibrium between Fe\,{\sc i} and Fe\,{\sc ii}. To be more precise,
in order to calculate $\sigma_{\rm mod}$, we studied the abundance changes
for two different {\em consistent} models (T$_{\rm eff}$=7750\,K, $\log g$=0.6)
and (T$_{\rm eff}$=8250\,K, $\log g$=1.5), together with a change of the
microturbulent velocity of $\xi_t$=2\,km\,s$^{-1}$. The total
uncertainty on the [el/Fe] abundances $\sigma_{\rm tot}$ can be found in
Table \ref{tab:abndcsboth}, and is the quadratic sum of the uncertainty on the
mean due to line-to-line scatter, the uncertainty induced by the model, and the
uncertainty on the Fe abundance:
\begin{displaymath}
\sigma_{\rm{tot}} = \sqrt{(\frac{\sigma_{\rm{ltl}}}{\sqrt{\rm{N}_{\rm{el}}}})^2 + (\sigma_{\rm{mod}})^2 + (\frac{\sigma_{\rm{Fe}}}{\sqrt{\rm{N}_{\rm{Fe}}}})^2
}
\end{displaymath}
If less than 5 lines were available, a line-to-line scatter of 0.2\,dex was
applied.


\section{Diffuse Interstellar Bands}\label{sect:diffuseisbnds}
Diffuse Interstellar Bands (DIBs) are broad absorption lines of interstellar
origin that are seen in the spectra of reddened objects. The carriers of these
DIBs are still not known, but polycyclic aromatic hydrocarbons (PAHs) are
amongst the most probable candidate carriers. Due to the severe mass loss in the
preceding AGB phase, post-AGB stars are often enshrouded by carbon-rich
circumstellar dust, causing severe reddening. Therefore, post-AGB stars are
ideal testlabs to search for possible {\em circumstellar} DIBs. If, on the other
hand, the DIBs are detected to be interstellar, a rough division can be made
between the interstellar and circumstellar component of the total reddening
towards the post-AGB star, since some DIBs correlate quite well with the
interstellar reddening.

\begin{figure}
\resizebox*{\hsize}{!}{\includegraphics{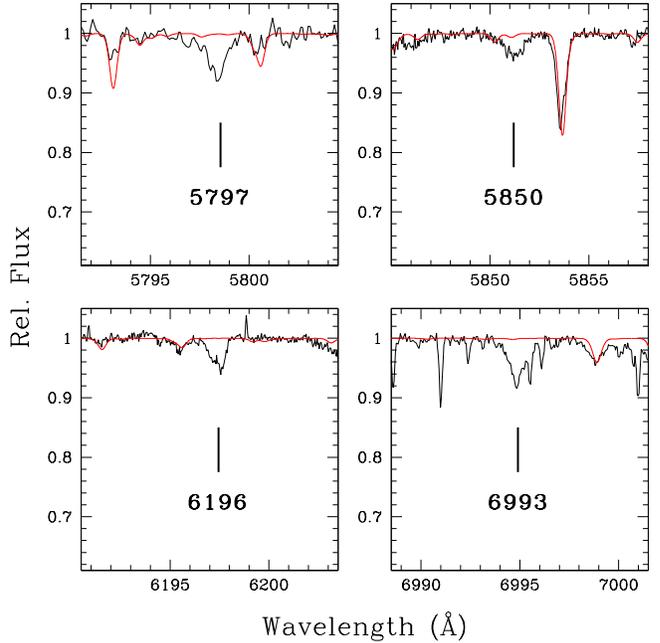}}
\caption{Four clear diffuse interstellar bands in the spectrum of
\object{IRAS\,14325-6428}. A spectral synthesis is overplotted in red, to
facilitate the DIB-detection.}\label{fig:fourdbs}
\end{figure}

\subsection{DIBs in \object{IRAS\,14325-6428}}
During our analysis of \object{IRAS\,14325-6428}, we realised that strong
DIBs are indeed present in this spectrum. We initiated a systematic search for
DIBs. First, we selected eight well-known and widely studied DIBs from the list
in \citet{herbig95}. For each of these DIBs, we made a spectral synthesis in
the vicinity of the DIB wavelength, based on the abundances found in our
abundance analysis (Table \ref{tab:abndcsboth}). The DIB can then easily be
identified as the spectral feature that is not fitted. Four DIBs are shown in
Fig. \ref{fig:fourdbs}. The studied DIBs in the spectrum of
\object{IRAS\,14325-6428} are listed in Table \ref{tab:dibtbl}. The columns
of this table represent: rest wavelength of the DIB; the observed wavelength
in \object{IRAS\,14325-6428}; the heliocentric radial velocity of the DIB; the
measured equivalent width; the intrinsic strength of the DIB as seen in the DIB
standard \object{HD\,183143}; and the (interstellar) reddening derived from the
latter relation. The reference for columns (1) and (5) is \citet{herbig95}.

The heliocentric radial velocities of the DIBs show some scatter, but
they are definitely different from the heliocentric velocity of the star
($-$87 km\,s$^{-1}$) and show velocities very different from any
expected outflow of the circumstellar material. Therefore, the DIBs seen in the spectrum of
\object{IRAS\,14325-6428} are the imprint of an {\em interstellar} cloud
between the object and the observer. A rough estimate for the distance towards
the DIB producing cloud is obtained through the formula of \citet{lang80},
which describes the rotation of the Galactic Plane, yielding 1.6\,kpc.
Obviously, this distance is also a lower limit for the distance towards
\object{IRAS\,14325-6428}. The scatter in the DIB velocities could be caused
by different DIB producing clouds in the line of sight towards
\object{IRAS\,14325-6428}, but also the rest wavelengths of the DIBs are
difficult to quantify due to their complex fine structure
\citep[e.g.][]{galazutdinov03}.

For most DIBs, there is a correlation between its strength and the
reddening that is caused by the DIB producing cloud. The DIB standard that is
often used to quantify this relation, is \object{HD\,183143}
(E(B-V)\,$=$\,1.28). In Table \ref{tab:dibtbl}, the DIB strengths of this
standard, normalized to E(B-V)\,$=$\,1, are given in column 5, together with
the inferred reddening for the DIB producing cloud towards
\object{IRAS\,14325-6428} (column 6). A mean interstellar reddening of
E(B-V)\,$=$\,0.6 is obtained. If we combine this with the {\em total} reddening
towards \object{IRAS\,14325-6428} of E(B-V)$_{\rm tot}$\,$=$\,1.1
(Table \ref{tab:sedsfinalrslts}), we can conclude that a significant fraction
(around 50\,\%) of the total reddening towards \object{IRAS\,14325-6428} is of
interstellar origin.

\subsection{DIBs in \object{IRAS\,08281-4850}}
Due to the lower resolution and the lower S/N ratio of the
\object{IRAS\,08281-4850} spectra, we were not able to do a similar study for
this source. A qualitative comparison of the strength of the strong DIB at
6196\,\AA\ indicates that \object{IRAS\,08281-4850} has certainly weaker DIBs
than \object{IRAS\,14325-6428}. The total reddening towards
\object{IRAS\,08281-4850} is, however, stronger than the one towards
\object{IRAS\,14325-6428}, implying that the circumstellar component of the
total reddening of \object{IRAS\,08281-4850} is probably large.

\begin{table}
\caption{Studied DIBs in the spectrum of \object{IRAS\,14325-6428}.}\label{tab:dibtbl}
\begin{tabular}{rrrrrr}
\hline\hline
DIB $\lambda$ & obs. $\lambda$ & V$_{\rm helio}$& EW & $\frac{\rm EW}{\rm E(B-V)}$& E(B-V)\\
(\AA)         & (\AA) & (km s$^{-1}$)  & (\AA) & {\scriptsize \object{HD\,183143}}  & {\scriptsize from DIB}\\
\hline
5780.45 & 5779.43 & $-$16.6 & 0.417 & 0.626 &  0.7\\
5796.98 & 5796.15 &  $-$6.3 & 0.087 & 0.186 &  0.5\\
5849.65 & 5849.57 &  $-$1.1 & 0.042 & 0.064 &  0.7\\
6195.95 & 6195.61 & $-$13.2 & 0.043 & 0.063 &  0.7\\
6283.86 & 6283.60 &  $-$9.2 & 1.043 & 1.520 &  0.7\\
6379.20 & 6378.95 &  $-$8.5 & 0.041 & 0.096 &  0.4\\
6613.62 & 6613.28 & $-$12.4 & 0.138 & 0.280 &  0.5\\
6993.07 & 6992.82 &  $-$7.7 & 0.079 & 0.142 &  0.6\\
\hline
mean\,($\pm\sigma)$ &\multicolumn{2}{r}{$-$9.4\,($\pm$4.8)} & & \multicolumn{2}{r}{0.6$\,(\pm$0.1)}\\
\hline
\end{tabular}
\end{table}


\section{Discussion}\label{sect:discssn}

\object{IRAS\,08281-4850} and \object{IRAS\,14325-6428} were selected on the
basis of their position in the IRAS colour-colour diagram \citep{pottasch88}
and the lack of free-free radio continuum emission
\citep{vandesteene93, vandesteene95}. 
Both objects remained poorly studied until now. Our chemical analysis presented
in this paper shows that both objects are among the hottest members of the
s-process enhanced post-AGB stars known to date and illustrates that the rich
IRAS legacy of objects in the transition between the AGB and the PN phase is
far from being harvested. Although the distance to the objects is not well
constrained, both objects do not show an extremely high luminosity
(Table \ref{tab:sedsfinalrslts}) and, given the sub-solar metallicity, both
objects represent the final evolutionary phase of a star with a low initial
mass (M$_{\rm i}$ $<$ 2 M$_{\odot}$). 

There is general agreement that the source of the neutrons for the
s-process in low and intermediate AGB stars is the ``$^{13}$C
source''. A long standing problem is the formation of the $^{13}$C
pocket itself. Stellar models using a standard treatment of mixing
cannot reproduce the $^{13}$C pocket at a level which is high enough
for the s-process to take place \citep{herwig05}. Until now, there is
no satisfying description for this phenomenon, and the different
modeling groups use different descriptions. Many models use some kind
of overshoot mechanism \citep[e.g.][]{herwig97}. Also differential
rotation has recently been studied as a possible mixing mechanism
driving the proton engulfment, but this rotationally induced mixing
alone cannot account for the formation of a large enough $^{13}$C
pocket \citep{herwig03}. Other groups avoid this formation problem by
assuming an ad-hoc $^{13}$C pocket in the He intershell
\citep[models by e.g.][]{gallino98} or an ad-hoc proton density profile
\citep[models by e.g.][]{goriely00}.
The observed spread in s-process efficiency (see Sect. \ref{sect:introdctn})
is then reproduced by a variable $^{13}$C pocket strength. Since the
$^{13}$C pocket is build upon primary synthesised $^{12}$C, the formation
is thought to be largely independent of the initial metallicity and a
richer s-process nucleosynthesis is expected for lower metallicity
stars, because more neutrons become available per iron-seed. In practise,
such a trend has not been found for post-AGB stars.

The very recent results of the population synthesis models by \citet{bonacic06}
and \citet{bonacic07} are very interesting in this context. Their
approach is to combine stellar population synthesis with a rapid stellar
evolution code including AGB nucleosynthesis and evolution. Interestingly,
their models do reproduce both the observed dichotomy
\citep[see Fig. 1d in][]{bonacic06}, and the observed spread. Contrary to
previous models, only a limited spread in the strength of the $^{13}$C pocket
is needed to reproduce the observed efficiency spread. With the new results
presented in this paper, we are able to add two important datapoints to
constrain this promising new generation of models. Indeed, the two stars
discussed here show a strong s-process enrichment, although they are only
mildly metal deficient. Particularly the results of \object{IRAS\,08281-4850}
are on the predicted boundaries of both the [hs/ls] index
\citep[Fig. 3 of][]{bonacic07} and the Zr overabundance
\citep[Fig. 8 of][]{bonacic07}. 



\section{Conclusion}
With the comprehensive abundance analysis presented in this paper, the objects
\object{IRAS\,08281-4850} and \object{IRAS\,14325-6428} join the group of
post-AGB stars with a clear post third dredge-up signature, since they do not
only show a clear carbon enhancement, but also a strong enrichment in s-process
elements. This enrichment is surprisingly strong with a high ratio of heavy
versus light s-process elements, despite the only mild metal deficiency of
both objects.  This provides additional evidence of the intrinsic
s-process efficiency spread in (post-)AGB stars. A systematic analysis
of post-AGB stars in the PNe locus of the IRAS colour-colour diagram
would be a very rewarding program to study the relation between the AGB
nucleosynthesis, the dredge-up efficiency and the overall stellar evolution of
the central star.

\begin{acknowledgements}
MR is grateful to Sara Regibo for the preliminary analysis of IRAS\,14325-6428
in an early stage of the paper. The Geneva staff is thanked for observation
time on the Euler telescope. MR acknowledges financial support from the Fund
for Scientific Research - Flanders (Belgium). PvH acknowledges support from the
Belgian Science Policy Office through grant MO/33/017.
\end{acknowledgements}

\bibliographystyle{aa}



\end{document}